\documentclass[12pt]{article}\usepackage[hyperfootnotes=false]{hyperref}
\usepackage{epsfig}
\usepackage{float}
\usepackage{empheq}
\usepackage{bbold}
\usepackage[utf8]{inputenc}
\usepackage{amsmath}

\usepackage{caption}

\usepackage{amsmath}
\usepackage{amssymb}
\usepackage{graphicx}
\newlength{\abstractwidth}
\renewcommand{\thefootnote}{\fnsymbol{footnote}}
\renewcommand{\thanks}[1]{\footnote{#1}} % Use this for footnotes
\newcommand{\starttext}{
\setcounter{footnote}{0}
\renewcommand{\thefootnote}{\arabic{footnote}}}

\newcommand{\be}{\begin{equation}}
\newcommand{\bea}{\begin{eqnarray}}
\newcommand{\eea}{\end{eqnarray}}
\newcommand{\beq}{\begin{equation}}
\newcommand{\ee}{\end{equation}}

\newcommand*\widefbox[1]{\fbox{\hspace{2em}#1\hspace{2em}}}

\def\eq{&=&}

\def\simleq{\; \raise0.3ex\hbox{$<$\kern-0.75em
\raise-1.1ex\hbox{$\sim$}}\; }
\def\simgeq{\; \raise0.3ex\hbox{$>$\kern-0.75em
\raise-1.1ex\hbox{$\sim$}}\; }

\def\bi{\begin{itemize}}
\def\ei{\end{itemize}}
\def\S{Schwarzschild}
\def\sc{\setcounter{equation}{0}}

\def\bsub{ \begin{subequations}
\begin{empheq}[box=\widefbox]{align}  }
\def\esub{ \end{empheq}
\end{subequations}}

\def\1{\(  \mathbb{1} \)}

 \def\lf{\left(}
    \def\rg{\right)}

  \def\bn{\bigskip \noindent}

 \def\bm{\begin{bmatrix}}
 \def\em{\end{bmatrix}}

\makeatletter
\g@addto@macro\normalsize{%
  \setlength\abovedisplayskip{10pt}
  \setlength\belowdisplayskip{20pt}
  \setlength\abovedisplayshortskip{10pt}
  \setlength\belowdisplayshortskip{20pt}
}
\makeatother

\usepackage{color}

\begin{document}

%%%%%%%%%%%%%%%%
%%%%%%%%%%%%%%%%

%%%%%%%%%%%%%%%%%%%
%%%%%%%%%%%%%%%%%%%%
  
\begin{titlepage}

\rightline{}
\bigskip
\bigskip\bigskip\bigskip\bigskip
\bigskip

\centerline{\Large \bf {  Black Hole-String Correspondence }} 

\bn

\bigskip
\begin{center}
\bf      Leonard Susskind  \rm

\bigskip
Stanford Institute for Theoretical Physics and Department of Physics, \\
Stanford University,
Stanford, CA 94305-4060, USA \\

and

Google, Mountain View, CA

%\vspace{1cm}
\end{center}

\bn

\begin{abstract}

I was asked to give a brief review of 
the  black hole-string correspondence \cite{Susskind:1993ws} as a warm-up for a longer SITP-group  discussion of a recent paper by Chen, Maldacena, and Witten \cite{Chen:2021dsw}. Here are my notes in written form.

\end{abstract}

\end{titlepage}

\starttext \baselineskip=17.63pt \setcounter{footnote}{0}

\Large

%\tableofcontents

\sc

\section{The Problem}

In 1993 I was invited by the Rutgers string-theory group to give a seminar on black holes and strings. I had thought about the relation between the two subjects a lot in the previous months. My picture was that the stretched horizon of a black hole is a thin layer of wiggly strings and that the entropy of the black hole is simply the entropy of those wiggles. I kept drawing the same picture which looked something like this.
\begin{figure}[H]
\begin{center}
\includegraphics[scale=.5]{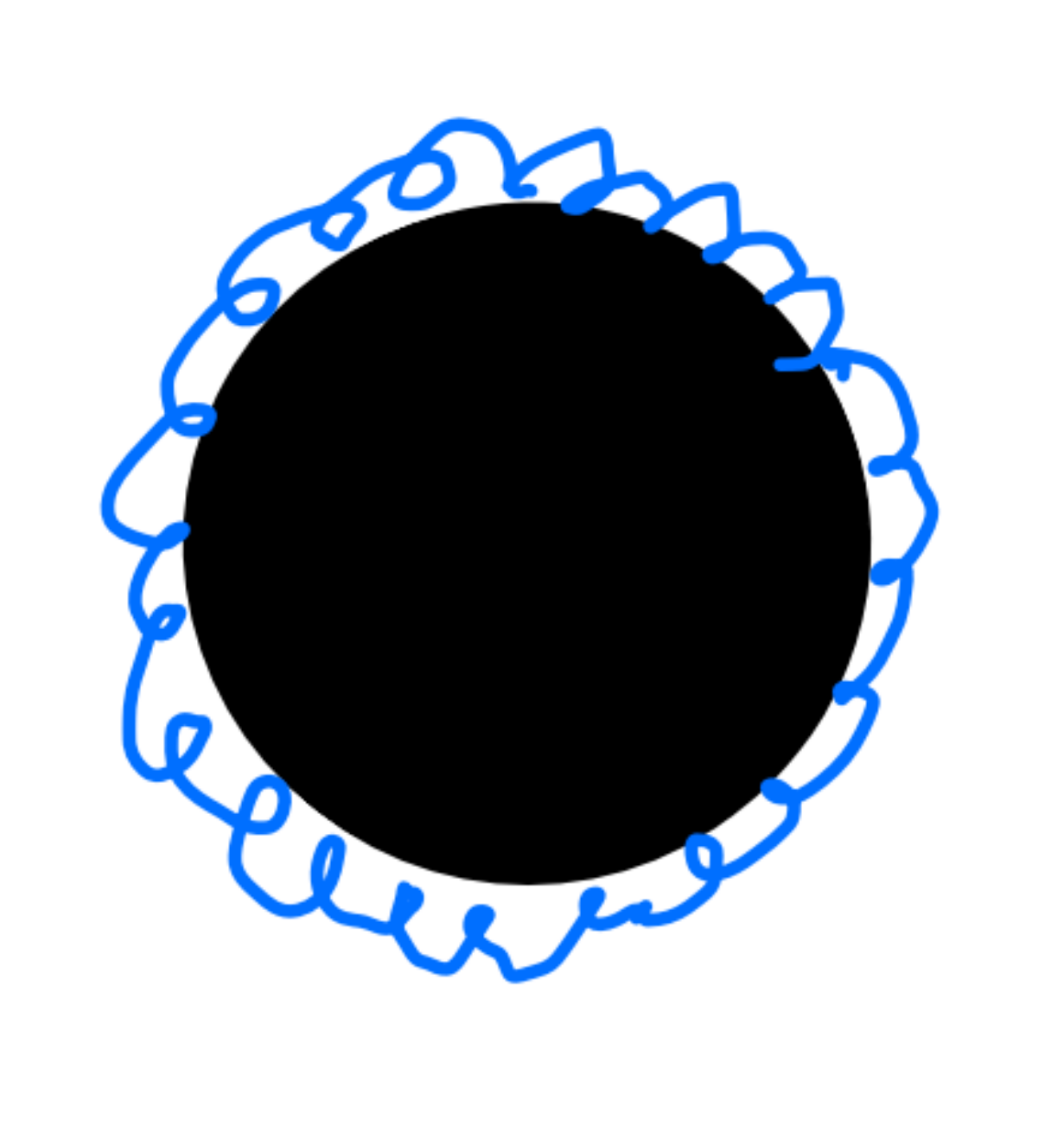}
\caption{}
\label{fuzz}
\end{center}
\end{figure}

What I really wanted was to use this picture to give a microscopic estimate of the entropy of a Schwarzschild black hole\footnote{At that time the idea that black hole entropy had a microscopic origin in some unitary quantum mechanics was largely dismissed by most relativists. } and show that it is proportional to the area of the horizon in Planck units. I thought this would make a fine seminar. The problem was that I had no idea how to do it. 

I spent all week before the seminar
trying to figure out a dynamical framework for the strings trapped by gravity just above the horizon, but I didn't see a way to do it. But at the last minute I had an idea:

\it
\bn
 Adiabatically vary the string coupling constant, or if you like, the background dilaton field until gravity gets so weak that it can no longer hold the string onto the horizon. When that happens the black hole should disappear and become a collection of almost free strings. If done slowly enough the entropy should not change during the course of the process (technically one would say that entropy is an adiabatic invariant), and we can calculate the original black hole entropy by using free string theory.

 \rm
  
  \bn

Now while it was clear that when gravity was switched off the black hole would have to become a collection of free strings, it was less clear how many strings would appear in the final state. One possibility was a large number of short strings, or maybe a mix of short strings and longer ones. The prospect of figuring out the quantum dynamics of the transition seemed very forbidding, but at some point I recalled a paper from the early days of string theory---I don't remember who wrote it---showing that the number of states of a single string of a given total energy is dominated by the states of a single long string. This meant that on statistical grounds, the final state of the adiabatic transition should be a single long string. This should  make the problem a lot easier.

\section{Some Facts}
Here are some facts from string theory and gravity that we will need. In what follows we will hold the string length-scale fixed\footnote{All equations are simplified by ignoring multiplicative factors of order unity. There would be no point in keeping these factors because a chain is no stronger than its weakest link, and there is one step in the argument that is only accurate to an order $1$ numerical factor.} and work in $(3+1)$-dimensions.
\subsection*{The Couplings}
The Newton constant and string coupling are related by,
\be 
G=g^2 l_s^2
\label{Geqg2l2}
\ee
\subsection*{The \S \ Radius}
\be 
R_{S} = MG = Mg^2 l_s^2 
\label{Srad}
\ee

\subsection*{Ratio of \S-radius to String Length}
From \eqref{Srad},
\be 
\frac{R_S}{l_s} = Mg^2 l_s
\label{ratio}
\ee

\subsection*{Black Hole Entropy}
\bea
S_{BH} &=& \frac{{\rm Area}}{4G} \cr \cr
\eq M^2 G \cr \cr
\eq M^2g^2{l_s}^2
\label{bhS}
\eea

\subsection*{Free String Entropy}
 It is a fact about strings that both the energy (mass) and entropy $S_s$ are proportional to the length of the string $L$ (measured along the string). On dimensional grounds,
\bea 
M \eq \frac{L}{l_s^2} \cr \cr
S_s \eq  \frac{L}{l_s}
\label{MandS}
\eea
implying
\be 
S_s=Ml_s.
\label{Ss}
\ee

\section{The Black Hole-String Transition}

\begin{figure}[H]
\begin{center}
\includegraphics[scale=.5]{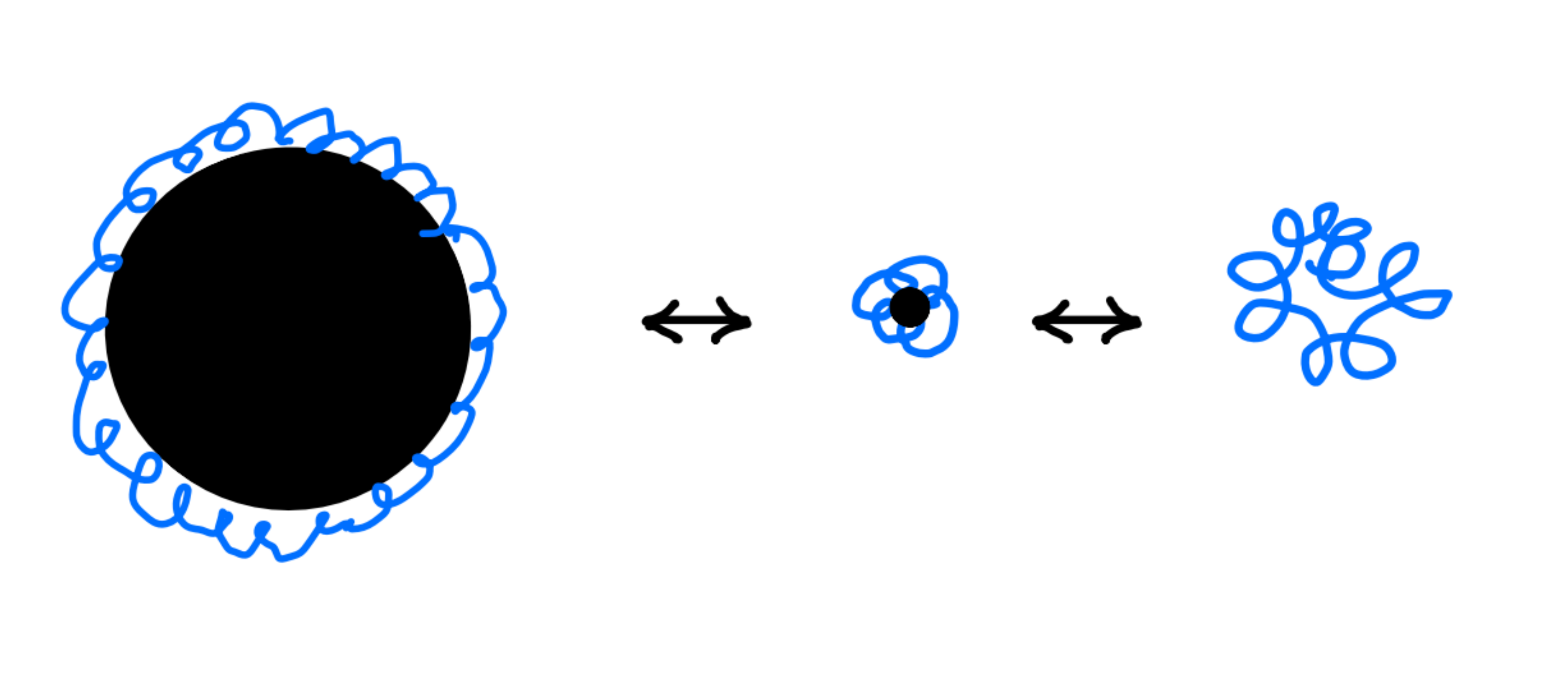}
\caption{Proceeding from left to right: A large black hole with a stringy stretched horizon, evolves, under adiabatic change of the coupling, to a black hole of string size, and then a single free string. }
\label{trans}
\end{center}
\end{figure}

Here is what I imagined the transition from black hole to free string looks like. As the string coupling decreases the \S \ radius in string units decreases \eqref{ratio}. No matter what the initial mass $M_0$ and initial Newton constant $G_0,$ 
eventually the Schwarzschild radius reaches the string scale, i.e., the size of the typical wiggles. That's what happens in the second picture of figure \ref{trans}. 

The key assumption in 1993 was that any further decrease in the coupling would result in the black hole being replaced by a single string.  Later, the guess that the transition takes place when the \S \ radius reaches the string scale was put on  firm footing by Horowitz and Polchinski \cite{Horowitz:1997jc}.

\subsection*{The Transition Curve}

\begin{figure}[H]
\begin{center}
\includegraphics[scale=.4]{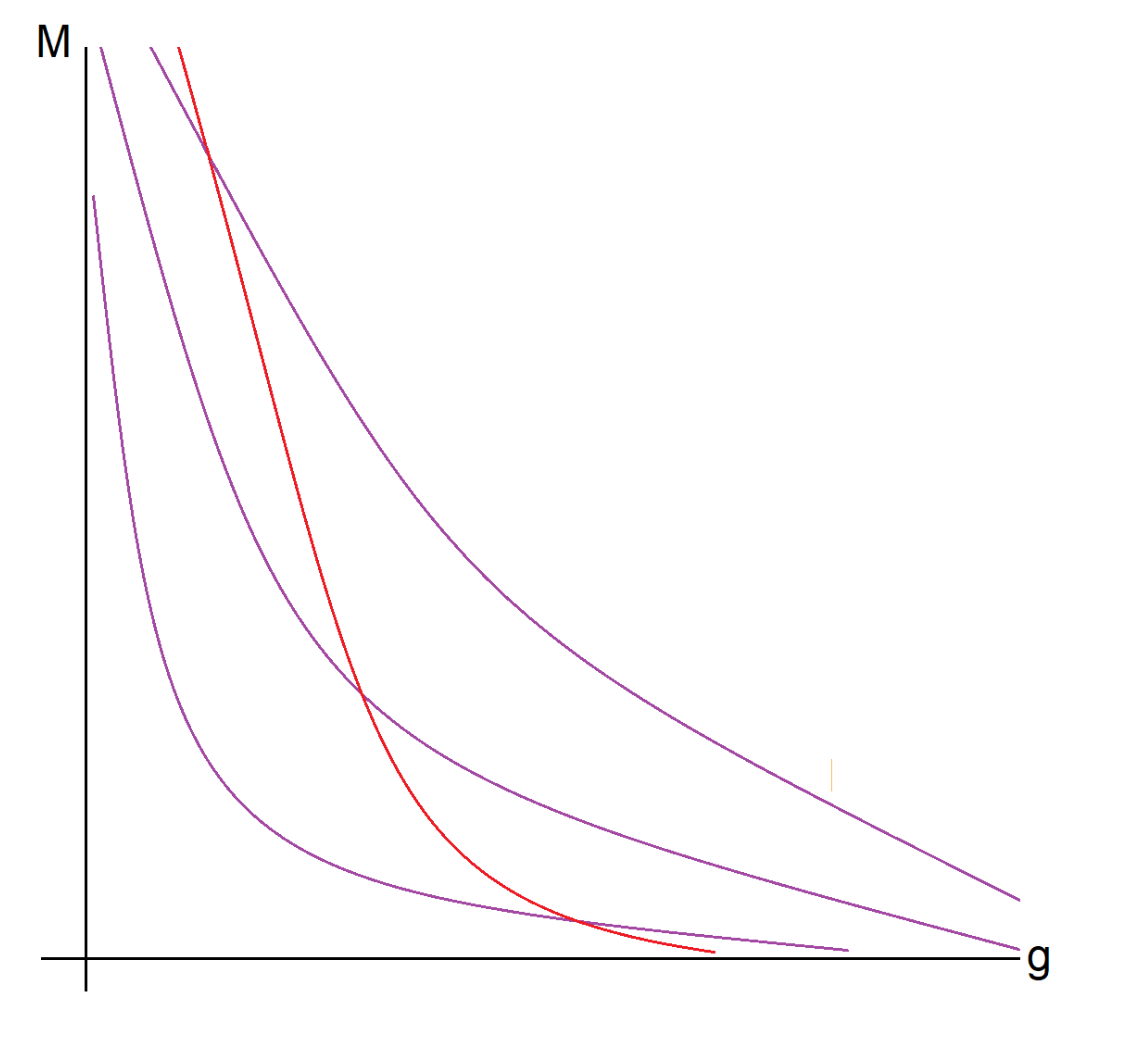}
\caption{The red curve is where the black hole-string transitions take place. The purple hyperbolas to the right of the red curve (the black hole region) are the lines of constant entropy, i.e., the adiabats.}
\label{graph}
\end{center}
\end{figure}

According to the key assumption the transition occurs at   
$\frac{R_S}{l_s} =1,$ or from \eqref{ratio},
\be 
M = \frac{1}{g^2 l_s}
\label{Meq1ovg2l}
\ee
I've shown this as the red curve in the diagram of figure \ref{graph}.

\bn
This transition curve defines the values ``matching points," i.e., the values  of $M$ and $g$ where the black hole and string descriptions coexist. 
\subsection*{The Adiabats}

Next let's construct the ``adiabats." Adiabats, as I learned when I was a mechanical engineering student, are the curves along which entropy is constant. Let's begin to the right of the transition curve---the black hole phase---where the entropy is given by \eqref{bhS}. The adiabats are clearly curves of constant $Mg,$ in other words the purple hyperbolas on the $M,g$ chart. 

What happens to the adiabats when they pass to the left of the transition curve? That's when the system becomes a free (or almost free) string. In that limit the entropy becomes independent of the coupling. The adiabats become flat.
\begin{figure}[H]
\begin{center}
 \includegraphics[scale=.3]{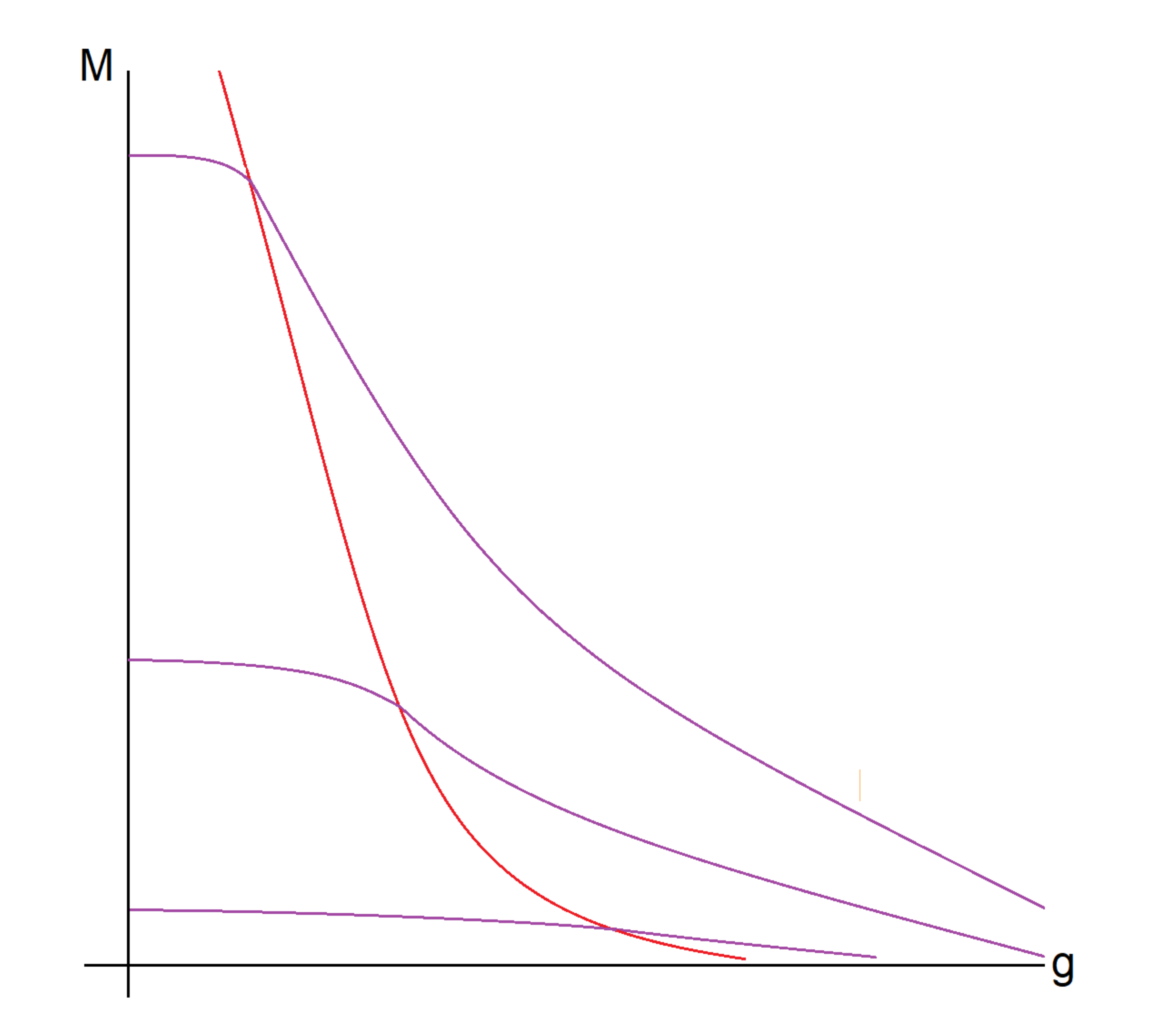}
\caption{The adiabats may be extended into the almost free string region where they flatten out. The reason is that when the coupling becomes very small the mass of a free string becomes insensitive to $g$. }
\label{{flatabs}}
\end{center}
\end{figure}
The simplest assumption was that the adiabats follow the hyperbolic trajectories until they intersect the transition curve and then flatten out. With that assumption the mass of a given adiabat when it hit $g=0$ is easy to compute---it's just the mass where the adiabat intersects the transition curve.

\subsection*{Tracking a Black Hole}

\begin{figure}[H]
\begin{center}
\includegraphics[scale=.3]{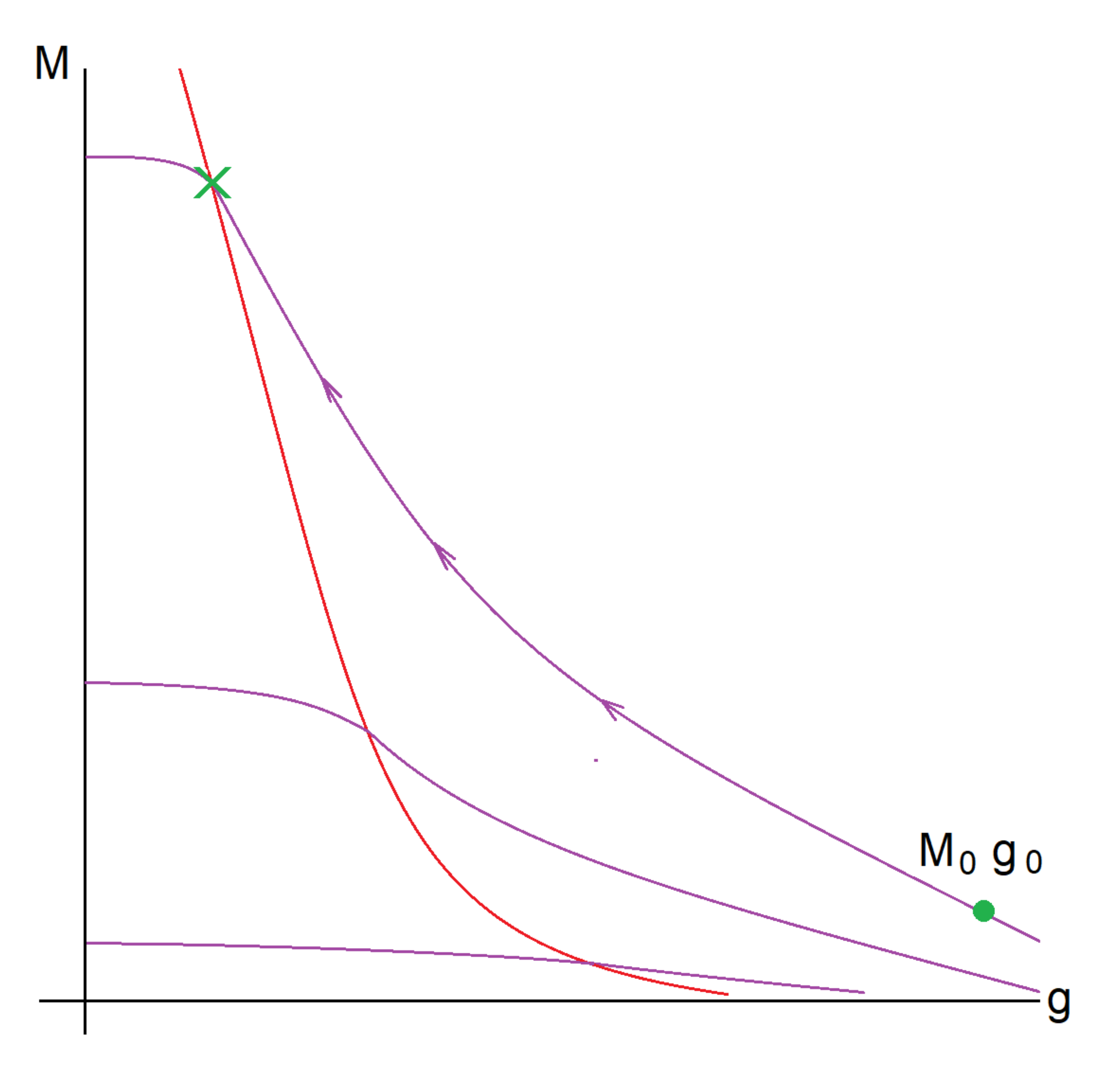}
\caption{Tracking a black hole: Start with a black hole of mass $M_0$ in a background with the string coupling being $g_0$ shown as a green dot. We may track it along an adiabat until it arrives at 
the transition point shown as a green cross.}
\label{track}
\end{center}
\end{figure}

Now to the point:  start with a black hole whose entropy we want to compute. The mass of the black hole is $M_0,$ the string coupling is $g_0$ and the Newton constant is $G_0 = g_0^2 l_s^2.$ I've plotted that point as a green dot in figure \ref{track}.

\bn
Next,  adiabatically decrease $g,$  and follow the black hole along its adiabat until it reaches the transition curve at the green cross. We want to know the mass and coupling constant at that point. Here are the equations for the transition curve and the adiabat:
\bea 
M \eq \frac{1}{g^2l_s}  \cr \cr
Mg &=& M_0g_0
\label{simult}
\eea
Solving them simultaneously gives the matching point (called the ``correspondence point" by Horowitz and Polchinski  \cite{Horowitz:1996nw}), 
\bea
g^2 &=& \frac{1}{M_0^2 G_0} \cr \cr
Ml_s &=& M_0^2 g_0^2 l_s^2
\label{solution}
\eea
The first equation for $g$ tells us that  at the correspondence  point it  is extremely small if the black hole mass is large in Planck units.  This is important in justifying the free string approximation.

\subsection*{The Result}
The second equation of \eqref{solution} is especially interesting. Using \eqref{Geqg2l2} and \eqref{Ss} it tells us that the entropy on the adiabat containing the point $(M_0, g_0)$ is given by
\be 
S = M_0^2 G_0.
\label{result}
\ee
This of course  is precisely what we hoped to get: the black hole entropy is given by the Hawking-Bekenstein formula written in the form  of the middle equation in \eqref{bhS}.

To summarize, what I did  is to match the black hole to a free string by adiabatically transporting the black hole parameters to the matching or correspondence point, and then calculate the entropy using free string theory. And it worked, giving the right relation between entropy and black hole mass.

At the time this was the first calculation to show that black hole entropy really does arise from the counting of quantum states.

\section{Entropy and Area}
So far I have not even mentioned the area of the horizon. Can we see that the entropy is related to the area by matching the area of a string (to be defined) to the entropy at the correspondence point? With the right interpretation we can. I'll give a very short intuitive explanation.

Let's recall a very general fact about black holes: the area of the horizon is exactly the zero-energy absorption cross section for a massless  scalar particle incident on a  black hole. Even away from zero energy the absorption cross section is proportional to the classical horizon area but with an order $1$ coefficient that varies modestly with energy.

If we ignore the long-range Newtonian elastic scattering (which leads to infinite cross section) then by the optical theorem the absorption cross section (and therefore the horizon area)
 is proportional to the imaginary part of the forward scattering amplitude. We can use this relationship as a definition of the area and extrapolate it to the correspondence point where we can compute it using string perturbation theory. This was done in  \cite{Halyo:1996xe}.  In this note I'll give a crude but effective way of estimating the result.

Depict the excited string as a closed random walk on a lattice in the $x,y$ plane. 
\begin{figure}[H]
\begin{center}
\includegraphics[scale=.5]{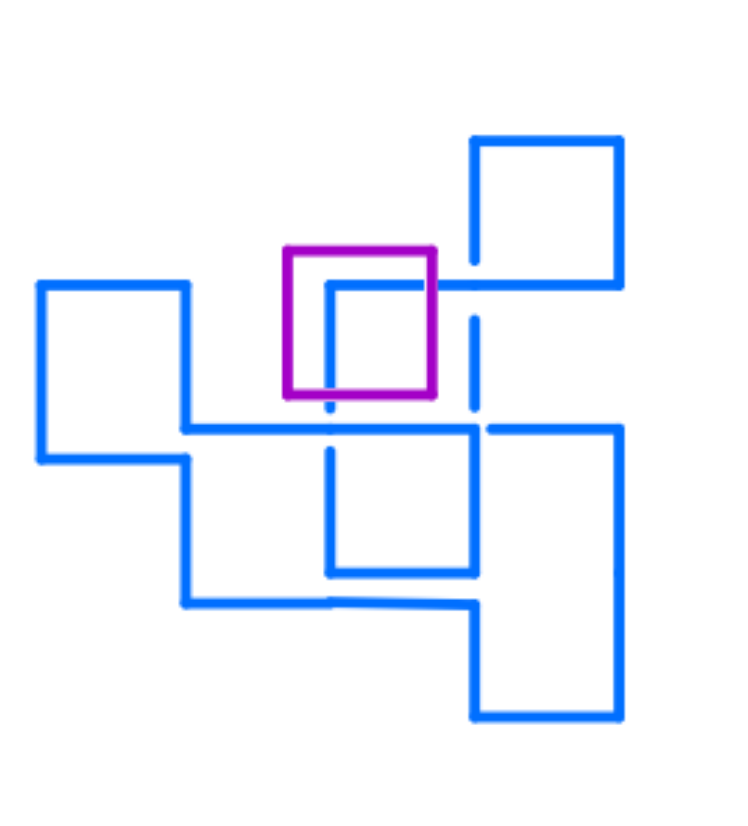}
\caption{A crude theory of the interaction between a highly excited string (blue) and a massless scalar string (purple).  }
\label{square}
\end{center}
\end{figure}
Each link has a length $l_s$ and the total length of the string is $L$. By  a standard argument the total mass of the string is,
\be 
M=\frac{L}{l_s^2}.
\label{Mfrac}
\ee

Now imagine a scalar particle represented by a small string of length $\sim 1.$ I'll draw it as a purple square.
The scalar particle moves along the $z$ axis, perpendicular to the $x,y$ plane. In figure \ref{square} the setup is illustrated with the excited string in blue and the scalar particle in purple.

Since at the correspondence point the coupling $g$ is very small we can assume that cross section  is just the sum of the cross sections for the scalar to collide with the individual links of the excited string. 
 
 The individual cross sections are obviously of order $g^2l_s^2.$ The factor $l_s^2$ must be there for dimensional reasons. The factor $g^2$ represents the  strength of the coupling.
 It follows that the total cross section is,
\be 
\sigma =\lf \frac{L}{l_s} \rg \lf g^2l_s^2 \rg.
\label{Sfrac}.
\ee

Now using \eqref{Geqg2l2} and  \eqref{Mfrac} we can write the cross section as,
\be 
\sigma =Ml_s G.
\label{sigone}
\ee
But according to \eqref{Ss}  $Ml_s$ is nothing but the entropy of the string, so \eqref{sigone} becomes,
$$ 
\sigma =S G,
$$
or, dividing by $G$ and identifying the cross section with the area of the horizon at the correspondence point, the result is just the Bekenstein relation,
\be 
S\sim A/G.
\ee

This may seem far from a rigorous demonstration that the   cross section is related to the entropy in the right way, but perturbative string theory allows a rigorous calculation of the absorption cross section. The calculation was carried out in \cite{Halyo:1996xe} and gives the same answer.

\section{Limits of the Method}
The method I used in 1993 was not up to the task of computing the numerical coefficient in the entropy-mass relation of the entropy-area relation. The main obstacle was the lack of detailed knowledge of how the mass of the system evolved over the transition region. In crossing the red line in figure \ref{track} the adiabat might jump one way or the other which would introduce a multiplicative uncertainty in the final outcome.   What was needed was a quantitative approach to the details of the transition. The rough arguments I gave were not sufficient for this purpose, so the precise coefficient of $1/4$ in the Bekenstein-Hawking formula was out of reach. 

One approach to the problem was immediately suggested by Vafa right after seminar in Rutgers. Vafa pointed out that if we applied similar reasoning to supersymmetric extremal BPS black holes we could be sure that the adiabats are exactly flat. That approach took a few years to work, primarily because there was no good example until Strominger and Vafa cooked up the D1-D5 system. That famously gave the factor of $1/4$ but only for extremal black holes.

Shortly after, Horowitz and Polchinski used the same method that I outlined to successfully estimate the entropy of a wide variety of non-extremal string theory black holes, with and without charge or angular momentum, and in various dimensions \cite{Horowitz:1996nw}, but again, the method was too crude to produce the factor of $1/4.$  The reason was the same: lack of a precise theory of the transition region. This led Horowitz and Polchinski to attempt  to build a dynamical theory of the transition \cite{Horowitz:1997jc}.

I won't describe their theory here except to say that it added an ingredient to string dynamics that had been previously left out. The ingredient was the Newtonian gravitational attraction between different parts of the long excited string. The thermal fluctuations of the string (which tend to spread it out in space) were counteracted by gravitational attraction that tended to pull the string together. All of this was done in a largely classical description of the string.

Taking account of these competing effects Horowitz and Polchinski gave a better  account of the transition, good enough to justify where the transition takes place, but still  not good enough to compute the numerical factor of $1/4$ with any precision.

It was the HP theory \cite{Horowitz:1997jc} that was the subject of the Chen, Maldacena Witten paper \cite{Chen:2021dsw} and the group discussion at SITP that I mentioned earlier.  As I understand the situation Chen, Maldacena Witten argued that the HP self-gravitating string is consistent in heterotic string theory, but there is some obstruction in type II string theory. The paper is technical but the bottom line is clear---there is still lots more to do to understand the entropy of generic black holes in string theory.
 I hope these notes will be useful to anyone who wants to pursue the subject further.

\section*{Acknowledgements}

Supported in part by NSF grant PHY-1720397.

\end{document}